\documentclass[prb,aps,showpacs,twocolumn,preprint numbers]{revtex4-1}

\usepackage{epsfig}
\usepackage{amsmath}
\usepackage{graphicx}% Include figure files
\usepackage{dcolumn}% Align table columns on decimal point
\usepackage{bm}% bold math
\usepackage{amssymb}
\usepackage{amsmath}
\usepackage{epsf}
\usepackage{subfigure}
\usepackage{epstopdf}
\usepackage{wrapfig}
\usepackage{pifont}
\usepackage{color}
\usepackage{wasysym}
\usepackage{slashed}
\usepackage{hyperref}
\usepackage{appendix}
\usepackage{subfigure}
\everymath{\displaystyle}

\begin{document}

\title{Infrared properties of cuprates in the pseudogap state: A study of 
Mitrovi\'{c}-Fiorucci and Sharapov-Carbotte scattering rates}

\author{Pankaj Bhalla}
\email{pankaj@prl.res.in}
\author{Navinder Singh}
\affiliation{Physical Research Laboratory, Navrangpura, Ahmedabad-380009 India.}

%%%%%%%%%%%%%%%%%%%%%%%%%%%%%%%%%%%%%%%%%%%%%%%%%%%%%%%%%%%%%%%%%%%%%%%%%%%%
%%%%%%%%%%%%%%%%%          Abstract                        %%%%%%%%%%%%%%%%%
%%%%%%%%%%%%%%%%%%%%%%%%%%%%%%%%%%%%%%%%%%%%%%%%%%%%%%%%%%%%%%%%%%%%%%%%%%%%

\begin{abstract}
The frequency dependent scattering rate of generalized Drude model contains 
important information on the electronic structure and on scattering mechanism. In 
the present investigation, we study the frequency dependent scattering rate of 
cuprates (Mitrovi\'{c}-Fiorucci/ Sharapov-Carbotte scattering rate)
in the pseudogap phase using the non-constant energy dependent 
Yang-Rice-Zhang (YRZ) density of states. First, with the energy dependent 
density of states, the scattering rate shows a depression at low energy
coming from the opening of the pseudogap. Second, the evolution of 
$1/\tau(\omega,T)$ with temperature shows the observed increase in scattering rate with 
temperature at lower frequencies and the temperature independence of 
$1/\tau(\omega)$ at higher frequencies. Third, the signature of the thresholds due 
to the boson density of states and to the electronic density of states are also 
observed. These signatures are qualitatively in accord with the experiments.
\end{abstract}
\date{\today}

\maketitle

%%%%%%%%%%%%%%%%%%%%%%%%%%%%%%%%%%%%%%%%%%%%%%%%%%%%%%%%%%%%%%%%%%%%%%%
%%%%                 Introduction                              %%%%%%%%
%%%%%%%%%%%%%%%%%%%%%%%%%%%%%%%%%%%%%%%%%%%%%%%%%%%%%%%%%%%%%%%%%%%%%%%

\section{Introduction}
\label{intro}
\begin{figure}
\centering
\hspace{0cm}
\subfigure[noonleline][]
{\label{pseudo1}\hspace*{-1cm}\includegraphics[height=4cm,width=5cm]{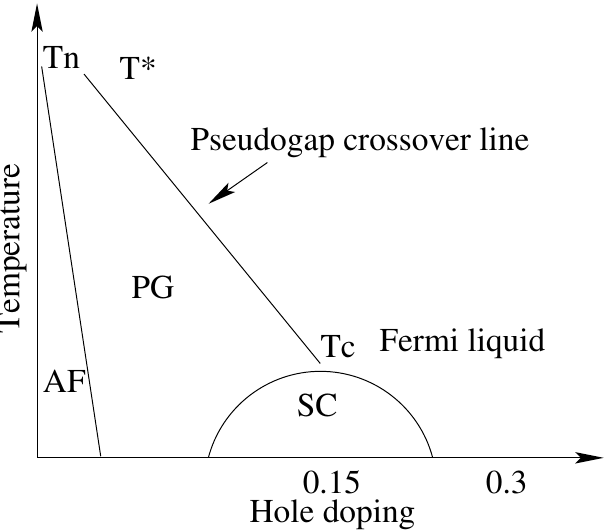}}
\hspace{1cm}
\subfigure[noonleline][]
{\label{cond1}\includegraphics[height=3cm,width=4cm]{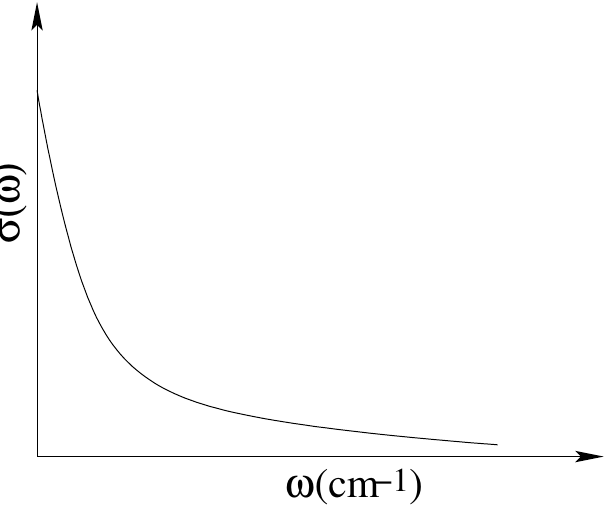}}
\hspace{0cm}
\subfigure[noonleline][]
{\label{cond2}\includegraphics[height=3cm,width=4cm]{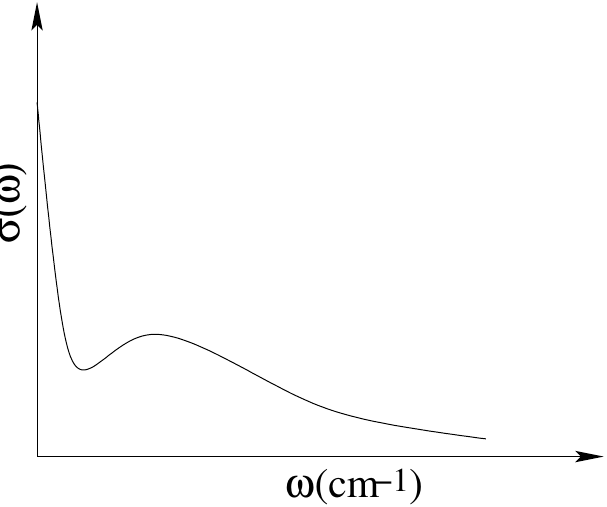}}
\caption{(a) The phase diagram of cuprates showing the three different regions 
with the hole doping and temperature variation. The used symbols are 
antiferromagnetic (AF), superconducting (SC), pseudogap (PG), Neel's 
temperature ($T_{n}$), Critical temperature ($T_{c}$), (b) the variation of the 
conductivity at temperature $T>T^{*}$ where there is no pseudogap and (c) 
corresponds to the behaviour of conductivity at $T<T^{*}$ where there is a pseudogap. There 
is a Drude part at lower frequencies, but there is small hump in conductivity at 
infrared part which gives the signature of the presence of the pseudogap.}
\end{figure}

High Temperature Superconductors such as cuprates \cite{bednortz_86} are strongly 
correlated systems\cite{anderson_87} having electron-electron interaction energy 
much greater than the electronic kinetic energy on the low doping side of the 
phase diagram and the opposite on the high doping side. In the intermediate 
regime of optimal
doping both are of comparable magnitude. This interplay of strong 
electron-electron interaction energy with their kinetic energy leads to the 
interesting phases as a function of temperature and doping. A typical phase 
diagram of the cuprates is shown in Fig.~\ref{pseudo1}. It has three different 
phases, antiferromagnetic (AF) phase, superconducting (SC) phase and the 
pseudogap (PG) phase. The present investigation deals with the pseudogap state, 
where physical properties of cuprates show anomalous behavior. These features 
are observed at the temperature $T<T^{*}$, where $T^{*}$ is the temperature 
below which the pseudogap is observed. Several attempts have been made to 
explain the physics of this region, but still there are some open questions, for 
example, the microscopic origin of the pseudogap state? What is the relationship 
(if any) of the pseudogap and superconducting gap? etc. The pseudogap is seen in 
many different experimental probes like Nuclear Magnetic Resonance (NMR), knight 
shift, neutron scattering, Angle Resolved Photoemission Spectroscopy (ARPES), 
tunneling experiments, etc. A clear manifestation of the pseudogap is also seen 
in infrared spectroscopy. In infrared spectroscopy, the reflectance of a given 
sample (at a given temperature and doping) is measured and then by 
Kramers-Kronig (KK) transformations, the infrared/optical conductivity 
$\sigma(\omega)$ is calculated. Below a temperature $T^{*}$, a depression is 
observed in the mid-infrared region of the conductivity which is a signature of 
the pseudogap formation\cite{timusk_03} as shown schematically in Fig. \ref{cond1} and \ref{cond2}. A more direct manifestation is observed in 
the frequency dependent scattering rate $1/\tau(\omega,T)$ of Generalized Drude 
Model (GDM) calculated from the experimental conductivity. This manifestation occurs
in the form of a depression below some characteristic frequency in $1/\tau(\omega,T)$. The 
representation of conductivity through frequency dependent scattering rate and 
mass enhancement factor is based on the work of Mori and Allen\cite{mori_65a,allen_71}, which is a generalization of the standard Drude model, when the 
electron-phonon coupling is important. The GDM can also be derived from Langevin 
equation with time dependent friction and it is of general validity\cite{kubo_66}. 
Allen's model gives a simple expression for the scattering rate in terms of 
the electron-phonon spectral function at zero temperature\cite{allen_71}. This is further 
extended for finite temperature by Shulga, Dolgov and Maksimov\cite{shulga_91}. But 
both of these formalisms are under the assumption of the constant Electronic 
Density Of States (EDOS) at the Fermi energy. Mitrovi\'{c} and Fiorucci\cite{mitrovic_85} 
have given a relation for non-constant EDOS at zero temperature that has been 
generalized recently by Sharapov and Carbotte\cite{sharapov_05} for finite temperatures. 

Much work\cite{dai_12,puchkov_96} has been done to explain the behaviour of 
$1/\tau(\omega,T)$ and the presence of pseudogap in cuprates. Puchkov et. 
al.\cite{puchkov_96} showed that when temperature $T$ is less than $T^{*}$ (the 
pseudogap temperature), absorption decreases at infrared frequencies which is a 
signature of pseudogap formation. Due to this decreased absorption, the 
in-plane conductivity shows the transfer of spectral weight from the low frequency 
Drude part to the mid-infrared part (see Fig.$~$\ref{cond1} and \ref{cond2}). 
Timusk et. al.\cite{timusk_03} also explained these experimental results and showed 
that at low frequencies and at temperature below $T^{*}$, the amplitude of the 
scattering rate is suppressed more and the mass enhancement factor increases. 
But no such type of signature has been observed at higher frequencies. The 
analysis of these results showed that there are two thresholds\cite{timusk_03} occurring in 
the generalized Drude scattering rate at around 600 cm$ ^{-1}$ and 1000 cm$^{-1}$. The main experimental features of $1/\tau(\omega,T)$ can be summed as:
\begin{enumerate}
\item
Depression formation at lower frequencies due to the opening of the pseudogap (refer Fig. \ref{schm1}).
\item
Two threshold scenario of $1/\tau(\omega,T)$ (refer Fig.\ref{schm}).
\item
Temperature dependence of $1/\tau(\omega,T)$ at low frequencies and temperature 
independence at high frequencies (refer Fig.\ref{schm2}).
\end{enumerate}

\textit{In the present investigation, we show that the above experimental features of 
$1/\tau(\omega,T)$ can be understood using the Mitrovi\'{c}-Fiorucci/Sharapov-Carbotte 
formulation of the scattering rate by taking into account the pseudogap 
formation in electronic density of states}. For the pseudogap EDOS, we use the YRZ 
phenomenological model\cite{yang_06,rice_12} and show that it describes the above 
mentioned experimental features of $1/\tau(\omega,T)$ very well. Thus, we report 
another success of the YRZ model.

The paper has been organized as follows. With the introductory background as given above
in section 1, we discuss the theoretical approaches to formulate the 
scattering rate and the YRZ ansatz in section 2. The numerical results and 
its analysis are given in section 3. In the last section, discussion and conclusion are presented.

%%%%%%%%%%%%%%%%%%%%%%%%%%%%%%%%%%%%%%%%%%%%%%%%%%%%%%%%%%%%%%%%%%%%
%%%%%            Theoretical approaches                        %%%%%
%%%%%%%%%%%%%%%%%%%%%%%%%%%%%%%%%%%%%%%%%%%%%%%%%%%%%%%%%%%%%%%%%%%%

\section{Theoretical formalism for scattering rate}
\label{sec:1}
Applying the Kinetic theory of gases to metals, Paul Drude \cite{drude_1900} formulated 
the formula for electrical conductivity:

\begin{equation}
\sigma(\omega) = \dfrac{\omega_{p}^2}{4\pi} \dfrac{1}{1/\tau - i\omega},
\end{equation}
where $\omega_p^2 = \frac{4\pi n e^2}{m}$, is the plasma frequency, $n$ is the 
free carrier density, $m$ is the carrier mass and $1/\tau$ is the scattering 
rate (the momentum relaxation rate). This formula can be derived from the momentum 
relaxation equation: $\frac{d\textbf{p}}{dt} = \frac{-\textbf{p}}{\tau} - 
e\textbf{E}$\cite{ashcroft_76}. The Drude model is valid when electrons are 
treated as free and only impurity scattering is important (momentum relaxation 
coming from impurity scattering). But if electron-electron or electron-phonon 
scattering is important, this formula breaks down. With the developments  in 
non-equilibrium statistical mechanics, especially by Kubo and Mori\cite{mori_65a,kubo_57,mori_65}, it 
became possible to treat non-equilibrium problems and interactions (perturbatively!) and to arrive at the 
generalized Drude model (see Allen\cite{allen_71}, Holstein\cite{holstein_64}, Gotze 
and Wolfe\cite{gotze_72}):

\begin{equation}
\sigma(\omega, T) = \dfrac{\omega_{p}^2}{4\pi} \dfrac{1}{1/\tau(\omega, T) - i\omega 
(1+\lambda(\omega, T))},
\end{equation}
where $\lambda(\omega, T)$ is the frequency dependent optical mass enhancement 
factor and $1/\tau(\omega, T)$ is the frequency dependent scattering 
rate\cite{sharapov_05}. On comparing the real and imaginary parts, one can obtain 
the frequency dependent scattering rate as:

\begin{equation}
\dfrac{1}{\tau(\omega, T)} = \dfrac{\omega_p^2}{4\pi} 
\mbox{Re}\left(\dfrac{1}{\sigma(\omega, T)}\right),
\end{equation}
and the mass enhancement factor $\lambda(\omega, T)$,

\begin{equation}
1+\lambda(\omega, T) = - \dfrac{\omega_p^2}{4 \pi} \dfrac{1}{\omega} 
\mbox{Im}\left(\dfrac{1}{\sigma(\omega, T)}\right).
\end{equation}

The plasma frequency $\omega_p$ can be calculated using the sum rule analysis: 
$\int_0^\infty d\omega \sigma(\omega,T) = \dfrac{\omega_p^2}{8}$. From the experimental 
optical conductivity $\sigma(\omega,T)$, $1/\tau(\omega,T)$ and $\lambda(\omega,T)$ can 
also be extracted.\\

Now, we would like to discuss the microscopic picture of $1/\tau(\omega,T)$. 
Analytically, Allen\cite{allen_71} has formulated the simple expression for 
$1/\tau(\omega)$ for metals by taking the electron-phonon interaction into 
account. It relates the $1/\tau(\omega)$ with electron-phonon spectral function 
$\alpha^2F(\omega)$:

\begin{equation}
\dfrac{1}{\tau(\omega)} = \dfrac{2 \pi}{\omega} \int_0^\infty d\Omega (\omega - 
\Omega) \alpha^2F(\Omega).
\end{equation}

The above zero temperature formalism of Allen was generalized by Shulga et. 
al.\cite{shulga_91} for finite temperature:

\begin{eqnarray}\nonumber
\dfrac{1}{\tau(\omega, T)} &=& \dfrac{\pi}{\omega} \int_0^\infty d\Omega I^2\chi(\Omega)
 \left[ 2 \omega \coth\left( \dfrac{\Omega}{2T}\right) \right. \\ \nonumber
&& \left. -(\omega + \Omega) \coth\left(\dfrac{\omega + \Omega}{2T}\right)\right. \\
&& \left.  + (\omega - \Omega) \coth\left(\dfrac{\omega - \Omega}{2T}\right) \right]
\label{sh}
\end{eqnarray}

\begin{figure}
\begin{center}
\includegraphics[height=5cm,width=8cm]{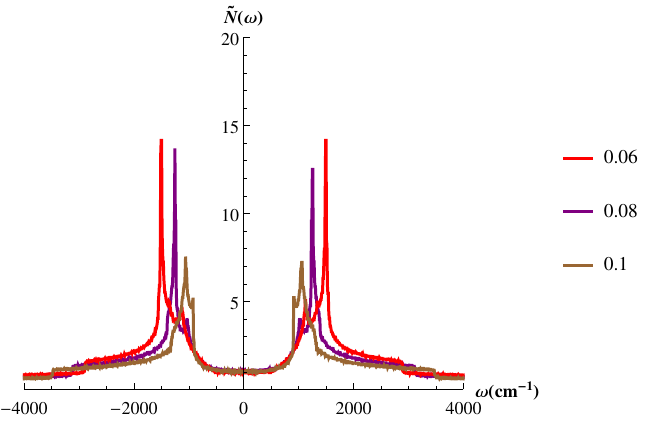}
\caption{The form of electronic density of states $\tilde{N}(\omega)$ using the 
YRZ model at different dopings such as 0.06, 0.08 and 0.10. Here, the 
hopping parameter that have been used is $t_{0}$=3600 cm$^{-1}$ (0.45eV). This 
plot shows the reduction of the magnitude of the pseudogap with the increase of 
the doping concentration.}
\label{dos}
\end{center}
\end{figure}

But with an important change: the electron-phonon spectral function 
$\alpha^2F(\Omega)$ was replaced by electron-boson spectral function 
$I^2\chi(\Omega)$\cite{sharapov_05}, to include the general case of scattering of 
electrons by boson degrees of freedom (other than phonons). Here the boson 
degrees of freedom could be bosons of electronic origin or the spin 
fluctuations\cite{carbotte_11}. 
With the limit $T\rightarrow 0$ in Eq. \ref{sh}, one can obtain the Allen's result. 
Most importantly, the above formulae are not valid for the pseudogap phase of 
the cuprates. Because the above formulae assume the constant electronic density 
of states (EDOS) near the Fermi level. In cuprates, the electronic density of 
states modifies \textit{significantly} at $T \leq T^{*}$ --the signature of pseudogap-- 
as has been confirmed by many other probes like ARPES, etc. Taking a non-constant 
electronic density of states into account, Mitrovi\'{c} and Fiorucci\cite{mitrovic_85} (in the context of metals) gave a relation for $T=0K$: 
\begin{equation}
\frac{1}{\tau(\omega)} = \dfrac{2 \pi}{\omega} \int_0^\omega d\Omega 
I^2\chi(\Omega) \int_0^{\omega - \Omega} d\omega' \tilde{N}(\omega'),
\label{mit}
\end{equation}
where $\tilde{N}(\omega) = \dfrac{N(\omega) + N(- \omega)}{2 N(0)}$, is the 
normalized and  symmetrized density of states\cite{mitrovic_85}. To include the 
temperature effects in this picture, recently, Sharapov and 
Carbotte \cite{sharapov_05} has given an important general expression:
\begin{eqnarray} \nonumber
\frac{1}{\tau(\omega, T)} &=& \frac{\pi}{\omega} \int_0^\infty d\Omega 
I^2\chi(\Omega) \int_{-\infty}^\infty d\omega' \tilde{N}(\omega'-\Omega)\\ \nonumber
&& \times \left[n_B(\Omega) + 1 - f(\omega'-\Omega)\right] \\ 
&& \times \left[f(\omega'-\omega) - f(\omega'+\omega)\right],
\end{eqnarray}
where $n_B(\omega)=\frac{1}{\exp(\beta \omega)-1}$ and $n_F(\omega)=\frac{1}{\exp(\beta 
\omega)+1}$ are the Boson and Fermi distribution functions and $\beta = \frac{1}{k_B T}$, 
$k_B$ is the Boltzmann constant. Several models are discussed in literature for the form of 
the electronic density of states, for example, a step function, a flat density of states, 
a triangular form, etc\cite{sharapov_05,hwang_08,hwang_13}. These toy model forms of 
$\tilde{N}(\omega)$ are modeled to take care of the formation of the 
pseudogap, but are quite far from the reality, for example, the pseudogap is known 
to evolve with doping but these forms do not have doping dependence. A realistic 
model for EDOS is the Yang-Rice-Zhang (YRZ) phenomenological model. The details of 
this model are given in Yang et. al.\cite{yang_06,rice_12} and are briefly presented 
in the appendix. The electronic density of states from YRZ model (see appendix) is 
given by
\begin{equation}
N(\omega) = \sum_\textbf{k} A_{YRZ}(\textbf{k}, \omega).
\end{equation}
Here, $A_{YRZ}(\textbf{k},\omega) = 
\frac{-1}{\pi} \mbox{Im} [G_{YRZ}(\textbf{k},\omega+i\epsilon)]$ is the YRZ spectral function. 
The form of EDOS ($\tilde{N}(\omega)$) as calculated from YRZ model is shown in 
Fig.$~$\ref{dos} at three different dopings $x$=0.06, 0.08 and 0.1. The 
Fermi surface corresponds to $\omega=0$. For $\omega<0$, the states are filled and for 
$\omega>0$, one has empty states. We notice from Fig.$~$\ref{dos} that with 
decreasing doping, the gap (around $\omega=0$) widens. This is in accord with 
pseudogap crossover line in Fig.$~$\ref{pseudo1}.

%%%%%%%%%%%%%%%%%%%%%%%%%%%%%%%%%%%%%%%%%%%%%%%%%%%%%%%%%%%%%%%%%%%%%%%%%
%%%%                 Numerical Results and analysis                 %%%%%
%%%%%%%%%%%%%%%%%%%%%%%%%%%%%%%%%%%%%%%%%%%%%%%%%%%%%%%%%%%%%%%%%%%%%%%%%

\section{Numerical Results and analysis}
\label{sec:2}

\subsection{Experimental features of ab-plane $1/\tau(\omega,T)$ (qualitatively)}
\label{sec:3}
\begin{figure}
\centering
\hspace{0cm}
\subfigure[noonleline][]
{\label{schm1}\hspace*{-1cm}\includegraphics[height=4cm,width=5cm]{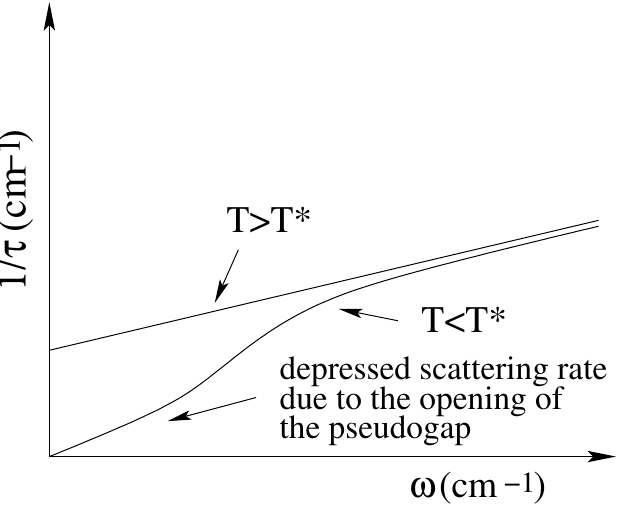}}
\hspace{1cm}
\subfigure[noonleline][]
{\label{schm}\includegraphics[height=3cm,width=4cm]{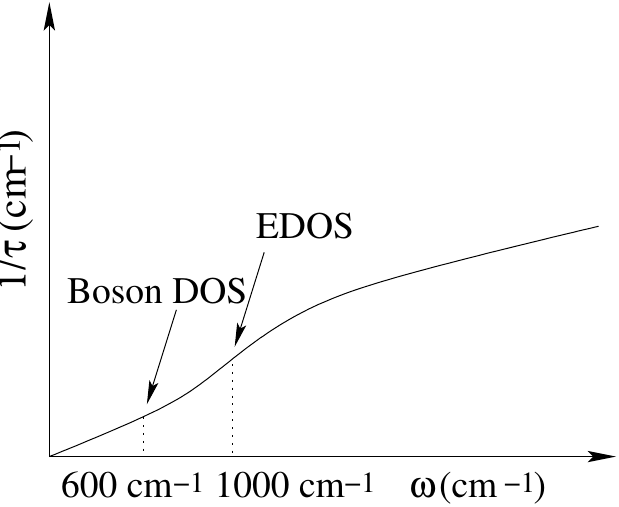}}
\hspace{0cm}
\subfigure[noonleline][]
{\label{schm2}\includegraphics[height=3cm,width=4cm]{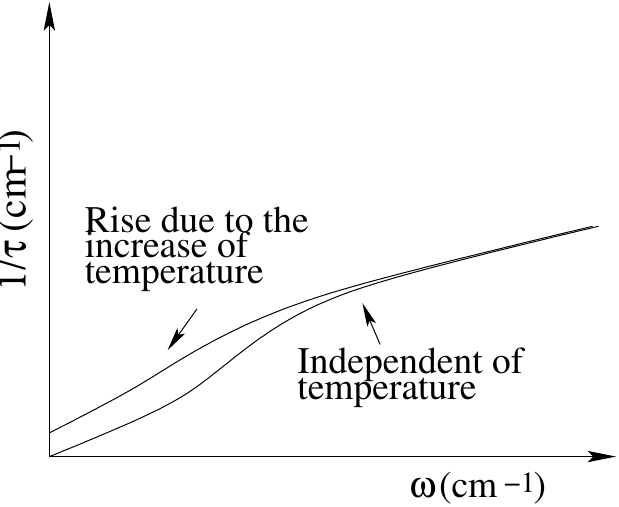}}
\caption{The schematic diagrams showing the qualitative behaviour of the 
experimentally observed scattering rate. (a) shows the depression picture due 
to the pseudogap formation at $T<T^{*}$ and no such feature at $T>T^{*}$. (b) 
The two thresholds scenario, one around 600 cm$^{-1}$ (possibly due to 
electron-boson scattering) and another around 1000 cm$^{-1}$ (due to the opening 
of pseudogap in EDOS). (c) Dependence of $1/\tau(\omega,T)$ on temperature at 
lower frequencies and its temperature independence at higher frequencies.}
\end{figure}

Experimentally, the scattering rate of the cuprates has been studied by many 
authors such as Lee et. al.\cite{lee_05}, Timusk et. al.\cite{timusk_03} and 
others\cite{carbotte_11,hwang_06,timusk_99,moon_14,liu_99}. A first important feature is the depression in 
the scattering rate at the low frequencies which is due to the opening of the 
pseudogap and occurs at $T<T^{*}$\cite{lee_05}. This is shown schematically in 
Fig.$~$\ref{schm1}. But no such type of depression has been seen at temperature $T>T^{*}$. 
 Secondly, Timusk proposes two depressions 
picture\cite{timusk_03}, one depression (at low frequency $\sim$ 600 cm$^{-1}$) is 
due to the electron-boson interaction and the other (at high frequency 
$\sim$ 1000 cm$^{-1}$) is due to the  electronic DOS (refer Fig.$~$\ref{schm}). 
A third and very interesting feature is the temperature evolution of the 
scattering rate (see Fig.$~$\ref{schm2}). In this case, there is an increase 
in the $1/\tau(\omega,T)$ with the increase of temperature only at low 
frequencies (roughly $\lesssim$ 1000 cm$^{-1}$) and at high frequencies (roughly 
$\gtrsim$ 4000 cm$^{-1}$), it does not show temperature dependence. 
In the next subsections, we analyze this observed behaviour of scattering rate 
using theoretical models.

\subsection{The Theoretical approach of Puchkov et. al.\cite{puchkov_96}}
\label{sec:4}
\begin{figure}
\centering
\subfigure[noonleline][]
{\label{shulga}\includegraphics[height=5cm,width=8cm]{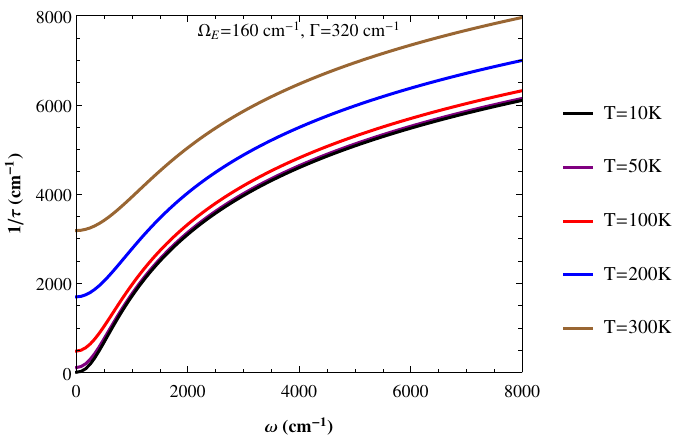}}
%\hspace{1in}
\subfigure[noonleline][]
{\label{puchkov}\includegraphics[height=5cm,width=8cm]{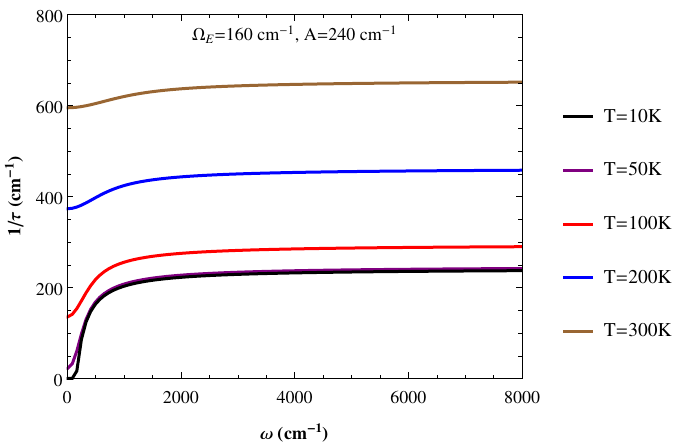}}
\caption{The behaviour of the scattering rate with the variation in temperature 
using the approach given by Puchkov. (a) using the Lorentzian model, (b) with 
the Einstein model. Here, the parameters used are shown inset of the graph.}
\end{figure}

Theoretically, Puchkov et al. \cite{puchkov_96} have investigated the behavior of 
the scattering rate at different temperatures. They consider the constant 
electronic density of states model (Allen's and Shulga-Dolgov-Maksimov's) and 
for the boson spectral function  ($I^{2}\chi(\Omega)$), they consider two models ($1$) the Lorentzian model
$ \Gamma\Omega/((\Omega - \Omega_{E})^2+(\Gamma)^2)$ with boson peak centered at 
$\Omega_{E}$ and width $\Gamma$, and ($2$) the single frequency $\Omega_E$ Einstein model 
$A\delta(\Omega-\Omega_{E})$ with $A$ as electron-boson coupling constant. With the Lorentzian model for 
$I^{2}\chi(\Omega)$, the observed features are shown in Fig.$~$\ref{shulga}. 
Here, we can see that at the low frequencies, there is a weak depression in the 
scattering rate at all temperatures. This depression comes from lower value of boson density of 
states at frequencies less than $\Omega_{E}$ that causes less scattering of 
electrons (thus depression in the scattering rate). But in the experiment, the 
depression (at lower frequencies) appears only at $T<T^{*}$\cite{lee_05}. No 
depression occurs when $T>T^{*}$. This experimental fact cannot be resolved 
using the boson only model of Puchkov et. al.\cite{puchkov_96}. It is hard to imagine 
that bosons suddenly becomes active at $T<T^{*}$ and inactive at $T>T^{*}$. We 
can also notice that with the increment in the temperature, the scattering rate 
increases at all frequencies in Puchkov et. al. approach whereas in experiment only the low frequency part is temperature dependent\cite{timusk_03}(refer Fig.\ref{schm2}). In 
Fig.$~$\ref{puchkov}, with the Einstein model, the $1/\tau(\omega,T)$ shows 
depression at lower frequencies and at lower temperature. But above about 1000 cm$^{-1}$, it saturates over a wide frequency scale. Contrary to this, it 
varies linearly in the high frequency range in experiment. The 
basic reason for this contradiction seen in Puchkov et. al's work is due to the 
assumption of the constant EDOS at the Fermi surface.

\subsection{Present approach: Mitrovi\'{c}-Fiorucci/Sharapov-Carbotte scattering 
rate with YRZ electronic DOS}
\label{sec:5}
To resolve these problems, we have performed numerical calculations for the 
scattering rate of cuprates using Mitrovi\'{c}-Fiorucci/Sharapov-Carbotte formalism 
with \textit{non-constant EDOS} and we see good agreement with experiment. In these 
calculations, we have used the realistic YRZ model\cite{yang_06,rice_12} for the EDOS. For the boson 
spectral function, we also consider two models ($1$) Lorentzian model $\Gamma 
\Omega/((\Omega - \Omega_{E})^2+(\Gamma)^2)$ and ($2$) Einstein model 
$A\delta(\Omega-\Omega_{E})$ as used in the previous investigations\cite{puchkov_96}. With 
this approach, we have computed the $1/\tau(\omega,T)$ in the underdoped region and at 
different temperature values. The used parameters are mentioned in the insets of the 
figures. But throughout the calculations, we have ignored the impurity contribution to the 
scattering rate (which contributes simple vertical constant shift of $1/\tau(\omega,T)$).

\begin{figure}
\begin{center}
\includegraphics[height=5cm,width=7cm]{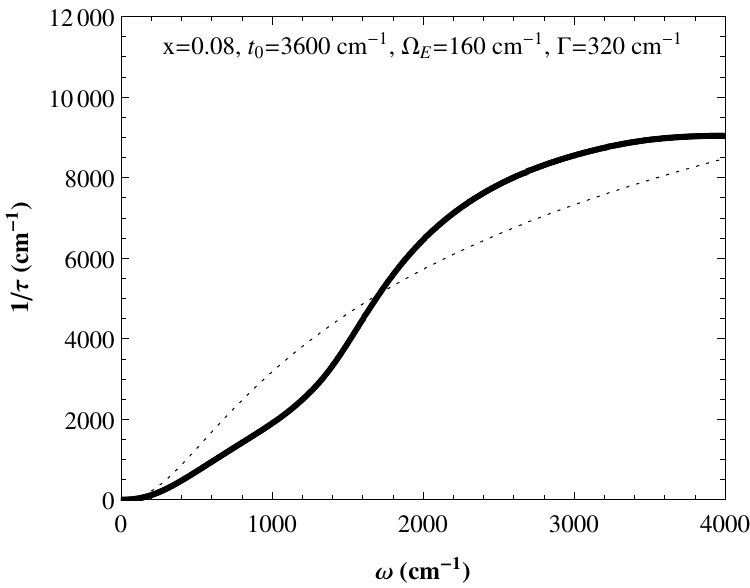}
\caption{Comparison plot for $1/\tau(\omega,T)$ against the frequency with (solid 
line) and without (dotted line) the consideration of the electronic density of 
states. Here, the parameters used are set as $\Omega_{E}$=160 cm$^{-1}$, 
$\Gamma$=320 cm$^{-1}$, x=0.08 and $t_{0}$=3600 cm$^{-1}$.}
\label{comp}
\end{center}
\end{figure}

As we mentioned before, a weak depression in Fig.$~$ \ref{shulga} in Puchkov et. 
al.'s analysis comes only due to electron-boson interaction and in the present 
approach (Fig.$~$\ref{comp}), the strong depression\cite{timusk_03,lee_05} comes predominately from the 
opening of the pseudogap which is also observed in experiment (see Fig.$~$3 in 
Ref. 23). \textit{From Fig.$~$\ref{comp}, it is clear that there is a 
sharper depression in the scattering rate with the consideration of 
non-constant EDOS as compared to the constant EDOS.} This strong depression feature 
clarifies the signature of the opening of the pseudogap, thus corroborates the experiment (refer Fig.$~$3 in 
Ref. 23). Another feature, the two thresholds picture as put forward by Timusk has also 
been noticed. 
First, around $\sim$ 160 cm$^{-1}$, where the boson density of states is maximum, the quasi-particles start to scatter more
and gives the signature of first threshold value. Further, in mid-infrared frequencies i.e. 
above $\sim$ 1200 cm$^{-1}$, the scattering rate produces another threshold due to 
the contribution of the increased electronic density of states above the pseudogap. But at higher frequencies, 
i.e. above 3000 cm$^{-1}$, due to the flattening of the EDOS (Fig.$~$\ref{dos}), the scattering rate flattens.

\begin{figure}
\begin{center}
\includegraphics[height=5cm,width=8cm]{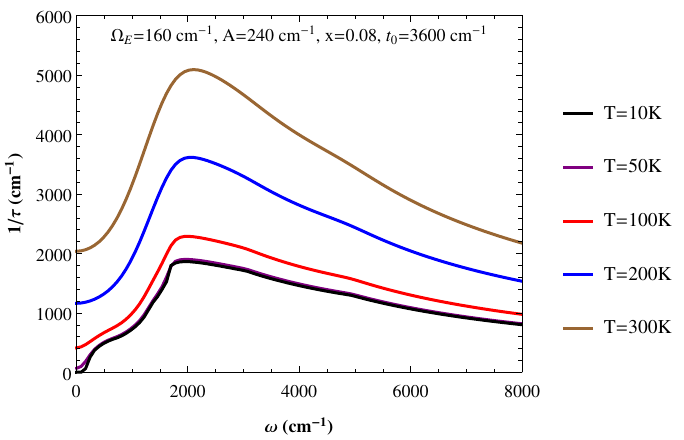}
\caption{With Einstein model, the temperature dependence character of the 
scattering rate at the doping concentration $x=0.08$. The scattering rate 
increases at all frequencies with the increase of the temperature.}
\label{einstein}
\end{center}
\end{figure}

\begin{figure}
\centering
\subfigure[noonleline][]
{\label{temp}
\includegraphics[height=5cm,width=8cm]{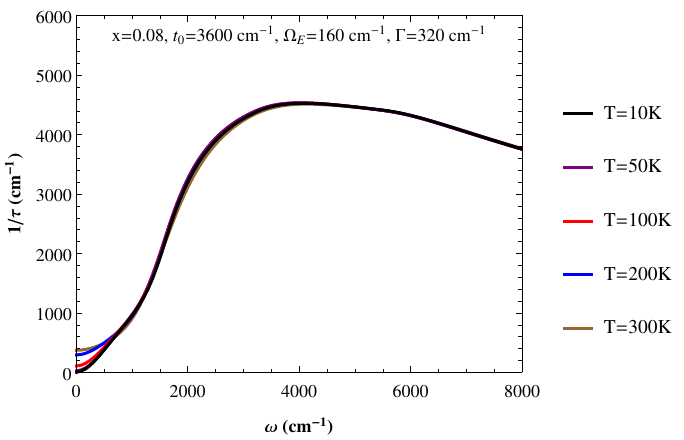}}
\hspace{1in}
\subfigure[noonleline][]
{\label{temp1}
\includegraphics[height=5cm,width=8cm]{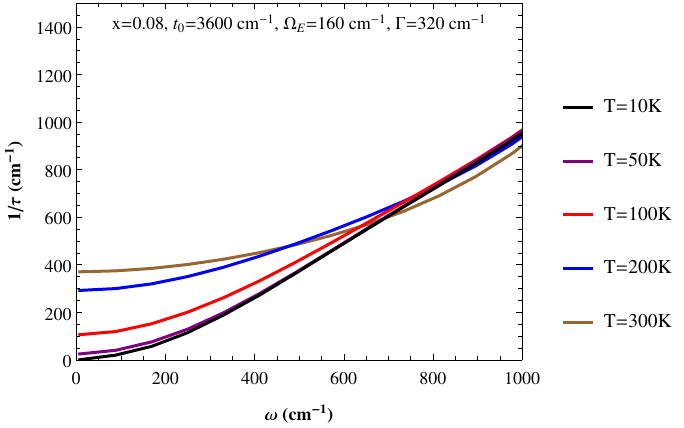}}
\caption{(a) The behaviour of the scattering rate at the doping $0.08$ with the 
change of the temperature. The scattering rate increases at the lower 
frequencies with the increase of the temperature and becomes temperature 
independent at higher frequencies. This is in accord with experiments. (b) The scattering rate on smaller frequency 
scale to see the clear view of the effect of temperature at lower frequency.}
\end{figure}

Now, we analyze the temperature dependence of the scattering rate at lower and 
higher frequency scale using both above mentioned models. First, using Einstein 
model, we can notice in the Fig.$~$\ref{einstein} that the scattering rate 
increases at all frequencies with the rise of temperature. But these results 
are not in accord with the experimental results. Because in this case, the boson 
mode has been set at a single frequency which is not an appropriate assumption. So, we have 
done the same calculations with the Lorentzian model. With this model, 
we can see in the Fig.$~$\ref{temp} that there is temperature dependence in 
$1/\tau(\omega,T)$ below a frequency $\sim$ 1000 cm$^{-1}$. But $1/\tau(\omega,T)$  is 
temperature independent at higher frequencies. These theoretically observed features are 
qualitatively in agreement with experiment. Thus we see that Sharapov-Carbotte scattering 
rate which include non - constant EDOS is in accord with experiments.

%%%%%%%%%%%%%%%%%%%%%%%%%%%%%%%%%%%%%%%%%%%%%%%%%%%%%%%%%%%%%%%%%%%%%%%%%%%%%%
%%%%%%%%%%%%%%%%%%%         Discussion and Conclusion     %%%%%%%%%%%%%%%%%%%%
%%%%%%%%%%%%%%%%%%%%%%%%%%%%%%%%%%%%%%%%%%%%%%%%%%%%%%%%%%%%%%%%%%%%%%%%%%%%%%

\section{Discussion and Conclusion}

One of the important messages of the present investigation is that with constant EDOS (in 
Puchkov et. al. approach as shown in Fig.\ref{shulga}), it is not possible to account for 
the temperature independence of $1/\tau(\omega,T)$ at higher frequency scale, whereas with 
non-constant EDOS (in Sharapov-Carbotte approach as shown in Fig.\ref{temp}), it is 
possible. This observation gives support to the Sharapov-Carbotte approach. The other points such as two thresholds picture and the strong depression 
formation in 
scattering rate has also been captured by the Sharapov-Carbotte formalism with the YRZ 
model as electronic density of states.
The problems that come across with the Puchkov et. al's approach has 
been sorted out. On comparing the results produced by Lorentzian and Einstein model, we 
have seen that the choice of Einstein model is inappropriate. Thus, the consideration of 
Lorentzian model for boson part and the YRZ model for electronic density of states in 
Sharapov-Carbotte scattering rate gives the qualitative picture of the experimental 
results.\\
But one important question remains, namely, in Fig.\ref{temp} we notice that the scattering 
rate decreases with increasing frequency above about 4000 cm$^{-1}$ whereas in experiment 
it increases linearly (refer Fig.\ref{schm}) with frequency. The answer to this question 
lies in the fact that YRZ model is a low energy model and at higher $\omega$, $N(\omega) 
\propto 1/\omega^{2}$ (from Eq. \ref{den}). This reduction in $N(\omega)$ at large $\omega$ 
(i.e. $\omega >> t_{0} \simeq$ 3600 cm$^{-1}$) leads to reduced scattering rate (see Eq. 
\ref{mit}). Another factor that causes the reduction in $1/\tau(\omega,T)$ is the toy model 
choice i.e. the Lorentzian model for Boson density of states. In the Lorentzian model, 
Boson density of states peaks at $\omega = \Omega_{E}$ and decays as power law at large 
$\omega$ ($\omega>>\Omega_{E}$). This also cause the reduction in scattering rate at higher 
$\omega$. In real materials, Boson density of states is much more sophisticated than the 
simple Lorentzian model\cite{hwang_14}. However, at lower energy scale ($\omega \lesssim 
t_{0}$) Sharapov-Carbotte formalism with Lorentzian model for Boson density of states 
\textit{qualitatively} reproduce the experimental features seen in $1/\tau(\omega,T)$.

\section*{Acknowledgement}
We are thankful to Prof. J. P. Carbotte for carefully reading the manuscript and suggesting 
many important corrections.

\appendix
\section{Yang-Rice-Zhang ansatz\cite{yang_06,rice_12}} 
\label{Appendix}
The Yang, Rice and Zhang have proposed the single particle propagator:

\begin{equation}
G(\textbf{k},\omega) = \frac{g_t(x)}{\omega - \xi(\textbf{k}) - 
\Sigma_{pg}(\textbf{k}, \omega)},
\label{green}
\end{equation}
where the self energy term is $\Sigma_{pg}(\textbf{k},\omega) = 
\dfrac{\Delta^{2}_{pg}}{\omega + \xi_{0}(\textbf{k})}$. Here $\omega$ is the 
frequency, $\textbf{k}$ is the momentum, $\Delta_{pg}= \frac{\Delta^{0}_{pg}}{2} 
(\cos(k_{x} a)-\cos(k_{y} a))$ is the pseudogap and $\xi(\textbf{k})$ is the 
band dispersion having hopping terms up to third nearest neighbor.
\begin{eqnarray}\nonumber
\xi(\textbf{k}) &=& -2t (\cos(k_x a) + \cos(k_y a))- 4t^{'} \cos(k_x a)\cos(k_y 
a) \\ \nonumber
&& - 2t^{''} (\cos(2k_x a) - \cos(2k_y a))) - \mu_p. \nonumber
\end{eqnarray}   
\begin{equation}
\xi_0(\textbf{k}) = -2t (\cos(k_x a) + \cos(k_y a)).
\end{equation}
The used parameters are: $\mu_p$ is the shift in the chemical potential, $t$, 
$t^{'}$, $t^{''}$, the hopping terms which are defined as $t = g_t(x) t_0 + 
3g_s(x) J \tilde{\chi}/8$, $t^{'}=g_t(x) t_{0}^{'}$ and $t^{''}=g_t(x) 
t_{0}^{''}$. And $\Delta_{pg}^{0} = 0.6 t_{0}\left(1-\dfrac{x}{0.2}\right)$. The 
Gutzwiller factors are $g_t(x)=\frac{2x}{1+x}$ and $g_s(x)=\frac{4}{(1+x)^2}$. 
The values of other parameters are $J=t_{0}/3$, $\chi=0.338$, $t_{0}^{'}=-0.3 
t_{0}$, $t_{0}^{''}=0.2 t_{0}$ and $x$ is the doping concentration\cite{yang_06,rice_12}.

The momentum averaged density of states in terms of spectral function\cite{sharapov_05} is defined as:
\begin{equation}
N(\omega) = \sum_\textbf{k} A(\textbf{k}, \omega),
\end{equation}
and $A(\textbf{k}, \omega) = \frac{-1}{\pi} \mbox{Im}\left[G_{YRZ} 
(\textbf{k},\omega+i\epsilon)\right]$. Therefore, with the help of the equation 
\ref{green}, the density of states can be written as:
\begin{equation}
N(\omega)=\frac{1}{\pi} \lim\limits_{\epsilon\rightarrow 0} \sum_\textbf{k} \frac{\epsilon g_{t}(x) 
(\omega+\xi_0(\textbf{k}))}{((\omega+\xi_0(\textbf{k})(\omega - \xi(\textbf{k}) 
- \Delta_{pg}^{2})^2+\epsilon^2(\omega+\xi_0(\textbf{k}))^2}.
\label{den}
\end{equation}

\end{document}